# Chapter 1

# Explainable AI, but explainable to whom?
## An explorative case study of xAI in Healthcare

Julie Gerlings[1,] Millie Søndergaard Jensen[2] and Arisa Shollo[1]

[1]Copenhagen Business School, Department of Digitalization, 31, Howitzvej 60, 2000 Frederiksberg, Denmark

e-mail: jge.digi@cbs.dk

[2]Aarhus University, Department of Linguistics, Cognitive Science and Semiotics, Langelandsgade 139, 8000 Aarhus C, Denmark

**Abstract.** Advances in AI technologies have resulted in superior levels of AI-based model performance. However, this has also led to a greater degree of model complexity, resulting in "black box" models. In response to the AI black box problem, the field of explainable AI (xAI) has emerged with the aim of providing explanations catered to human understanding, trust, and transparency. Yet, we still have a limited understanding of how xAI addresses the need for explainable AI in the context of healthcare. Our research explores the differing explanation needs amongst stakeholders during the development of an AI-system for classifying COVID-19 patients for the ICU. We demonstrate that there is a constellation of stakeholders who have different explanation needs, not just the "user." Further, the findings demonstrate how the need for xAI emerges through concerns associated with specific stakeholder groups i.e., the development team, subject matter experts, decision makers, and the audience. Our findings contribute to the expansion of xAI by highlighting that different stakeholders have different explanation needs. From a practical perspective, the study provides insights on how AI systems can be adjusted to support different stakeholders needs, ensuring better implementation and operation in a healthcare context.

## 1.1 Introduction

The adoption, use and diffusion of Artificial Intelligence (AI) technologies in healthcare comprise an area that currently receives vast amounts of attention from both researchers and people working in the industry. This increased attention is highly motivated by advances in machine learning (ML) technology, which have significantly improved the average performance of AI systems. ML-based systems are able to inductively learn rules from training data, typically consisting of input-output pairs (e.g., historical medical images and corresponding diagnoses). Once the rules are learned, they can be applied to make inferences about unseen data (e.g., medical images of a new patient). ML-based AI systems have enabled the automation and augmentation of sophisticated tasks previously performed exclusively by medical specialists. However, these systems are often built on non-transparent models where it is unclear how a model arrives at a given prediction. These systems are consequently often referred to as 'black box' models [1], [2]. Explainable AI (xAI) has emerged as a response to this 'black box' modelling; and the field of xAI seeks to provide explanations that accommodate human understanding and transparency [3]–[5].

In a healthcare context, existing literature on the role of xAI suggests that xAI is believed to enhance the trust of medical professionals interacting with AI systems. At the same time, rising legal and privacy aspects [6] are other driving factors in the development of xAI for the healthcare sector. Other areas where xAI is emphasized as being relevant in relation to AI in healthcare include accountability issues, reliability, justification, and risk reduction [7], [8]. These are all areas that are important to explore in order to secure successful AI implementation in healthcare. However, much of the research done so far is conceptual in nature, and there is currently limited knowledge on the need for xAI in applied healthcare contexts. In this chapter, we aim to gain further insights on this concept by exploring how the need for xAI arises during the development of AI applications and which issues xAI can help alleviate in an empirical setting. To investigate this, we conducted a case study and followed an AI startup during their development of an AI-based product for the healthcare sector. We guided our case study with the following research question:

*"How does the need for xAI emerge during the development of an AI application?"*

The empirical setting for our case study is a Nordic healthtech company specializing in medical imaging; the company is widely recognized as a startup with extremely strong professional competence in this area. Our study explores the development phase of an AI-based medical imaging product the company developed during the COVID-19 crisis; this product will hereafter be referred to as LungX. Based on a desire to utilize their competences to help alleviate some of the strain on the health services the company developed LungX to assist in automatic early evaluation of COVID-19 patients. One of the main challenges facing health services in relation to the COVID-19 crisis is predicting how the disease develops for each patient and thereby affects the resources of a given hospital [9]. Our investigations follow the development of LungX with a particular focus on identifying the ability of xAI to accommodate the needs of different stakeholders during the product life cycle.

The remainder of the chapter is structured as follows: the next section provides an overview of related literature and a detailed description of our methods and the empirical setting of our research. We then present the findings of the study before closing with a discussion on the theoretical and practical implications of our findings.

## 1.2 Related Work

As a starting point for answering our research question, we drew on prior work at the intersection of AI, xAI, and medical work with a primary focus on radiology. In the following section, we present relevant literature describing the drivers of AI adoption in healthcare, the emergence of xAI, and lastly, how both AI and xAI have been employed in the fight against the COVID-19 pandemic.

### 1.2.1 Adoption and use of AI in healthcare

AI-based algorithms are increasingly being developed for use in medical applications [2], [7], [10]. AI offers medical professionals the capacity to develop powerful, precise models capable of delivering individually tailored treatments backed up by aggregated and anonymized healthcare data [9], [11]. Hence, AI technologies create new opportunities for discovering new and improved treatment plans, facilitating early detection of diseases, and monitoring disease progression [12]. Further, these powerful technologies promise to solve problems of sparse resources by carrying out tasks currently operated by specialists such as radiologists, whose availability and mobility are limited [13], [14]. At the same time, they address the increasing need to keep up with the growing population and provide higher quality healthcare. However, the majority of these complex AI-based models remain in the development phase, never reaching production [9], [11], [12].

The increasing presence of complex AI models in healthcare has generated strong debates, as outcomes have varied [9], [15]. Models portrayed have failed to meet performance expectations, jeopardized critical decisions, and used discriminatory factors in determining outcomes [5], [13], [15]–[18]. Projects such as "Watson for Oncology" [19], an unnamed, broadly used algorithm to manage the health of large populations [20], Google Health [21] which can identify signs of diabetic retinopathy, or COMPAS [22] a decision support tool used in court houses to assess risk of recidivism, have all failed one way or another when put into test or even production. This has spawned a backlash against AI-based outcomes due to the potentially severe consequences they might have for humans making high-stakes medical decisions.

Improving the performance of machine learning models often requires increased complexity of the underlying models, larger datasets, and more computing power [4], [23], [24]. However, due to this heightened complexity, it becomes impossible to understand how these models work, how the data is processed, and how outcomes are generated [7], [25]–[27]. Hence, scholars have characterized such models as opaque or black box models [28]–[30]. In black box models metrics such as accuracy, precision and prediction speed take a front seat, hindering the ability of the general population to understand relevant outcomes, creating it an illiterate ability for the few. Their use in high-stakes decision-making, such as medical decision-making [2], [31]–[34] has led to increasing demands for transparent and explainable AI. Recent studies show that model transparency, interpretable results, and an understanding of clinical workflow are all necessary in order to ensure the correct use and adoption of these powerful models [2], [35], [36] in healthcare practices.

Apart from generating transparent or explainable models, it has also proven difficult to organize cross-disciplinary work between healthcare workers and data scientists to enable essential knowledge-sharing. Recently, scholars have addressed the issue of an 'AI Chasm' in the clinical research community [27], [36]–[38], identifying a gap in the literature between building a sound and scientifically correct model and using it in a real-world medical clinic. Though the EHR (electronic health record) has improved data volume in some parts of the world, both data quality and availability for training AI models are still prominent issues [11]. Lastly, regulatory obstacles like FDA approval and/or CE certification, which are typically necessary to obtain clinical trials, are slow processes that demand high performance results and documentation [36], [39]. These circumstances create a catch-22 situation where the production of higher-quality models and implementations is inhibited by restricted access to relevant data for training and achieving the desired performance scores and knowledge about how experts work.

### 1.2.2 Drivers for xAI

Many scientists suggest that xAI could ease and promote the adoption of AI in the medical domain; as explainability could accommodate understanding and trust, making stakeholders more willing to adopt a given AI application [6]–[8], [11], [40]. Studies on what kind of information is needed when a complex model is introduced in a decision-making context indicate that information provided by xAI frameworks may be of great relevance [2], [10], [35]. For example, Cai et al. [2] found that clinicians are interested in the local, case-specific reasoning behind a model decision as well as the global properties of the model. These information needs are similar to the information clinicians need when interacting with medical colleagues to discuss a patient case. Both the local and global model information could potentially be provided by xAI frameworks such as LIME, SHAP or PDP [41], [42]. In connection with this, Lebovitz [11] studied how an AI application aimed at diagnostic support in radiology was used in practice; the study found that the introduction of AI in the decision-making context introduced additional ambiguity into medical decision-making and caused their routine decision-making tasks to become nonroutine. The study emphasized that the lack of information about the model workings intensified the degree of ambiguity and that knowing more about how the algorithm was trained would have increased confidence in the results of the AI application. However, there remain very few studies investigating the actual information needed when introducing AI in algorithm-assisted decision-making in the medical sector. These studies motivated our research in xAI approaches and how they might be able to provide information to ease the understanding, adoption and implementation of complex AI-based models.

### 1.2.3 Emergence of xAI

The demand for transparency and explainable AI has emerged as a response to the increasing "black box" problem of AI. xAI refers to methods and techniques that seek to provide insights into the outcome of a ML model and present it in qualitative, understandable terms or visualizations to the stakeholders of the model [4], [6], [24], [43]. Moreover, [1] describe xAI as follows: *"Explainability is associated with the notion of explanation as an interface between humans and a decision maker that is, at the same time, both an accurate proxy of the decision maker and comprehensible to humans"*.

The terms explanation, transparency and interpretation all appear frequently in research dealing with the "black box" problem of AI, but the field has not yet reached a consistent, universal consensus on their meaning. This is partly due to the nascent state of the field and the fact that it spans multiple disciplines. xAI itself is the newest concept associated with human understanding of models and the efforts made in response to opaque deep neural networks [3], [44]. Explainability entails that a meaningful explanation can be formulated for a stakeholder [1]. Interpretability, on the other hand, entails the ability to assign subjective meaning to an object [44]. Interpretation occurs when a human uses their cognitive capabilities to form meaning from information. Transparency has a much more technical origin within the field of xAI. Here, 'transparency' is mainly used in computer science literature to describe different frameworks or methods to simplify complex models into binary models, rule generators or the like [1], [5], [42], [45]–[50].

When defining xAI products in relation to their utility for humans/model stakeholders, one key dilemma is the trade-off between complexity and interpretability. Oftentimes, scientists and engineers must choose between a more interpretable but less complex model and a less interpretable model that may offer a more accurate representation of reality. Another important trade-off arises in the form of accurate explanations versus comprehensible explanations; the more accurate the explanation, the more incomprehensible it will be for AI-illiterate stakeholders. Even though it may be possible to use xAI frameworks to accurately depict the model output or how the model produced the output, this description may still be incomprehensible to AI-illiterate stakeholders [24], [51], [52]. While these dilemmas have been acknowledged by scholars, there is a limited understanding of how they are dealt with in practice.

Social scientists have begun to address the request for more knowledge on the constitution of explanations — conceptual papers such as [46] have addressed the origins of explanations themselves, how we are biased in our explanation interpretation, and how explanations are phenomena that occur in the context of human interaction. Moreover, researchers from the socio-technical HCI-related fields [53], [54] problematize the way xAI research is approached by data scientists. Because xAI technology has its origins in data science, it was initially used as a tool for debugging and variable exploration. These studies implicitly adopt a universal conceptualization of explanations as useful for either developers or users. Further research is required to understand the constellation of stakeholders for AI models—the receivers of explanations—and the need for explanations they deem satisfactory. Moreover, scholars have similarly called for a more nuanced understanding of xAI and how it can satisfy different stakeholder needs by building more targeted explanations [7], [32], [55]–[57].

### 1.2.4 AI and xAI in the fight against the COVID-19 pandemic.

Before the WHO (World Health Organization) had even announced COVID-19 as a potential pandemic, AI-assisted and autonomous systems had succeeded in detecting and predicting the spread and severity of the pandemic [58]. Systems like BlueDot and HealthMap, which previously helped identify and detect SARS and the Zika virus (other coronaviruses), were now the first to identify unusual viral activity in Wuhan, China [59]–[62]. Algorithms utilizing big data and people's whereabouts both on- and offline helped warn the authorities about the impending crisis, helping public health officials respond in a fairly fast manner [61]. However, as we have seen, pandemics are extremely hard to combat; therefore, methods for identifying infected patients and disease severity are vital steps in this fight [18], [63].

The most common way to diagnose COVID-19 is by using the Reverse Transcription Polymerase Chain Reaction (RT-PCR), which is an expensive, time-consuming, resource-heavy, and complicated process. This has fueled the search for alternative ways of detecting the virus [18], [58], [63]. Recent research has shown promising results from methods using CT scans and lung X-rays to detect COVID-19 [64], [65]. However, the sparsity of radiologists who need to describe the images and give a diagnostics report have spurred the use of AI-based models to assist in this process.

Current literature identifies two main approaches that address COVID-19 in chest X-rays (CXR): a classification and detection problem, or a severity measure approach. The first approach is concerned with identifying COVID-19 *and* other lung-related pathologies (or anomaly detection, which tries to identify only COVID-19 infections from healthy CXRs). The classification and detection approaches have the same goal of identifying COVID-19, whereas the other approach is concerned with a severity measure for disease progression itself [66]. As the pandemic has evolved and the infrastructure for administering RT-PCR tests has greatly improved, the scope of AI-supported image scans has leaned towards disease progression and severity measures instead of disease diagnosis. Data used for training these deep neural networks (DNN) primarily consists of open-source CXR image repositories, which often contain labeled images of viral and bacterial pneumonia, fractures, and tuberculosis. However, image quality has proven suboptimal, as original X-ray images offer much higher quality and contain more helpful metadata in the DICOM (Digital Imaging and Communications in Medicine format than the images often available in a compressed format, such as JPEG [67]–[69].

A recent study [68] on how data processing and its variability in deep learning models affect explainability describes the fallacies of biased DNNs, which are not visible without some deeper insight into the models: *"...it is unclear if the good results are due to the actual capability of the system to extract information related to the pathology or due to the capabilities of the system to learn other aspects biasing and compromising the results."* [68]. Researchers use xAI to gain insight into the inner workings of their model and how it detects pneumonia in CXR. In particular, they have utilized the Gradient-weighted Class Activation Mapping (Grad-CAM) method [70], originally a debugging tool for DNN, to visualize the output of the model with a heatmap overlay. This visualizes the pixels of interest in the image and colors them in different shades according to their importance in relation to the output. This way, the visualization can be more easily validated by developers or radiologists to see if the correct pixels (variables) are marked by the model. Others have used the LIME framework [6] for similar purposes, helping radiologists by visualizing the model and identifying which features play a crucial role in distinguishing between COVID-19 patients and other patients [71]. Hryniewska et al. [15] point out that when a radiologist was shown the assumptions about data, models and explanations, many of the models created for COVID-19 were deemed incorrect. They have therefore created a useful checklist for building and optimizing future model performance and reliability.

The above studies show how xAI can be used to provide human-interpretable explanations for radiologists using the model for e.g., decision support or developers either debugging or testing the model's performance and diminishing biased data. This nascent research on the fight against COVID-19 appears to follow the same steps as the more general literature on healthcare xAI. Overall, we observe that current studies view xAI as advanced methods that provide universal explanations without taking into consideration the specific needs of different stakeholders.

In light of the above and the concept of explainability as socially constructed [46], we conducted a case study to examine how the need for xAI emerges and how the needs of different stakeholders are taken into account. Hence, this study aims to shed light on how xAI influences the development, adoption and use of AI-based products in healthcare settings.

### 1.3 Method

The empirical basis for this research is an illustrative interpretive case study that investigates the use of xAI in a healthcare setting. Case studies are particularly valuable for exploratory research where a thorough understanding of a phenomenon in a particular context is preferred [72], [73]. Further, the choice of the method depends on the nature of the question that is being investigated. Case studies are best suited for investigating 'how' questions [74], in this instance how the need for xAI emerges during the development of an AI-based healthcare project.

Against this background, we conducted a longitudinal case study in a growing Nordic start-up developing AI-based tools for decision support in healthcare. The start-up is currently working on two different products; whereof the one we investigated is aimed at detecting COVID-19 based on lung X-rays and providing an automated severity score. Two of the researchers collected empirical data over a period of six months, focusing on the development process of the product.

#### 1.3.1 Data Collection

The primary data source was semi-structured interviews with employees at the company, online workshops, and a collection of written documents produced by the company. Due to the circumstances of COVID-19, real-time observations were very limited, and most interviews and workshops were conducted via Microsoft Teams. Moreover, follow-up questions and updates from both sides were exchanged through email.

Interviews were conducted by two researchers, one leading the interview and the other following up on relevant points and remarks from the interviewee. Participants were informed of the purpose of the interview, the focus of our research, and the fact that the interviews would be anonymized. The interviews lasted an average of 50 minutes, and they were conducted in English. All the interviews were recorded and transcribed with the consent of the interviewees.

We conducted nine interviews with seven employees involved in the project, whereof two interviews were follow-up interviews with the CTO and a medical annotator (see Figure 1 for more details). We did not interview end-users such as radiologists, ICU clinicians, or patients, since LungX was not tested or implemented in hospitals during our research period. The interviews were based on an interview guide. Hence, we started with demographic and open-ended questions, followed by questions focusing on the history of the project, the interviewees' daily work with the project, data quality, use of xAI, model requirements, and future expectations. The interviews provided us with a thorough understanding of the project, including insights into both business and technical aspects of the product. While limited, the interviews were sufficient for the purposes of proof of concept.

Background information (including business proposal, PowerPoint presentations, project plans, grant application, a demonstration of the model and the user interface, user manual, requirement documentation and meeting minutes) was also collected as complementary material to the interview data.

| Role in Company: | Main responsibilities: | Interview themes |
|---|---|---|
| CEO | Developing the business case of product; responsible for business-related communication with clinics | Business aspect of the product, understanding of clinical operations, decisions for pursuing the case |
| CTO (2 interviews) | Designing the entire product (conceptual as well as technical parts); making research prototype | Technical perspective of the case, medical imaging and product details, reasoning behind decisions |
| Developer | Converting machine learning research code into quality-assured code for the product | Deeper insights into the product, including how the model is constructed |
| Developer | Fetching data from online sources and storing it for training; aligning technical aspects of LungX project with existing projects | Data quality and annotation process |
| Product Owner | Medical annotation of data; developing new features the product | Data quality, annotation process, validation |
| | should include (medical advisory role) | |
| **Medical annotator, (2 interviews)** | Medical annotation of data | Data quality, annotation process, validation |
| **Clinical Operations Officer** | Will be involved in planning the clinical validation of the product and day-to-day interactions with clinics; has not been very involved in early phases of product development | Project testing procedures, obstacles in product testing at hospitals, deep understanding of clinical processes at hospitals |

*Figure 1 Interviewee summary*

### 1.3.2 Data analysis

For the data analysis, researchers adopted an "insider-outsider" interpretive approach [75], [76] where initially, they established the "insider" perspective of how the need for xAI emerged in the AI-based project.

Once the "insider" understanding was formed, we engaged on a more abstract, theoretical level—the so-called "outsider" point of view—where the researchers created a link between the four dimensions [77]. In other words, we associated the "insider" point of view with the "outsider" point of view to merge our understanding of practice with our understanding of the existing literature.

The first and second author undertook the entirety of the field work, jointly developing an "insider" view of the process. The third author looked at the data after collection — having an "outsider" perspective on the phenomenon and the research site allowed the third author to provide new ways of theorizing and identify new patterns in the data which were discussed with the other authors. This method of data analysis revealed itself to be very fruitful since the first two authors could draw on their rich understanding of the data during the discussions of the identified patterns by agreeing or disagreeing with the third author, thus linking the "insider" view with the existing literature.

Our qualitative data analysis followed the approach of Gioia, Corley, & Hamilton [76]. We employed constant comparative techniques and open coding [78] in order to analyze our data, and our data analysis was an iterative process during which we discussed codes until we reached an agreement.

First, we moved all interview transcripts and background materials to NVIVO software to look for specific indicators of the practices where the need for explanations emerged in the project. These practices constitute our unit of analysis.

Next, we started by applying open coding on the interview transcripts to identify first-order concepts, staying close to the informant's original statements. We linked these original concepts together and formulated second-order themes (axial coding). Finally, we further grouped the themes together to come up with four dimensions through which the need for xAI manifested itself in the project. We illustrate our data structure in Figure 2, where we depict the concepts, themes and dimensions identified in the data analysis. As our coding of the data progressed, the first-order concepts emerged from quotes by interviewees. These were then grouped into second-order themes reflecting a higher-level concurrent theme structure in the coding. Lastly, these themes were aggregated into four dimensions of stakeholder's concerns emerging in the development of LungX: Development Team concerns, Subject Matter Expert (SME) needs and concerns, Decision-Maker needs and concerns, and Audience: (Patient) concerns. These concerns are constituted by the model's different stakeholders, articulating their own relationship to the model or their perception of other stakeholder groups in the development phase of LungX. Assessing the second-order themes, it is important to note that they are not necessarily mutually exclusive to the different dimensions but rather have the most influence on the depicted dimension. These stakeholder concerns are discussed in more detail in the findings section.

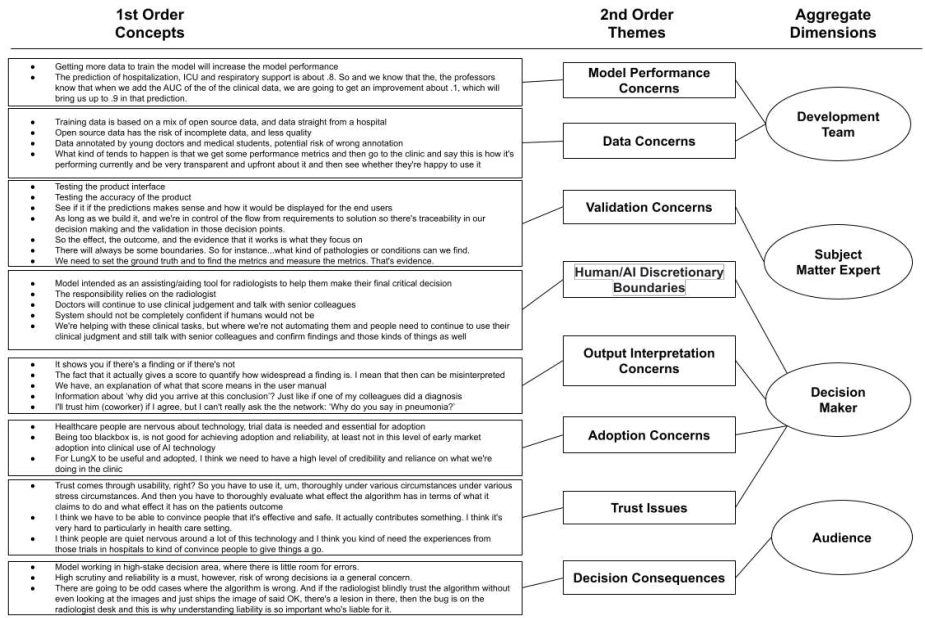

*Figure 2 Data Structure*

Going forward, we present the case setting followed by the findings from the data analysis and structure, thereafter a discussion of the findings, conclusion, and further research.

## 1.4 Case Setting

The empirical setting for our case study is a Nordic healthtech startup that specializes in medical imaging. The company was founded within the last five years with the vision of simplifying radiology and improving the patient journey through diagnostics. The five founders come from a mix of business and university backgrounds — four of them have extensive experience within medical imaging, while one of them has many years of experience with the commercial execution of life science projects. The company has approximately 20 full-time employees and a handful of part-time associates organized within the areas of business, research, development, operations, and regulatory affairs. When our research project started, the startup had one Class I CE-marked product on the market: a medical imaging solution able to triage patients and provide diagnostic support based on brain MRI scans. While it was not the subject of our focus, the existing brain product is worth mentioning because it came up in several interviews as some of the elements of LungX are inspired by the brain solution. Our research is centered around the development of LungX, a new solution based on lung X-rays.

The idea to develop LungX emerged based on a desire to alleviate strain on health services providers during the COVID-19 crisis by putting the company's medical imaging capabilities to use. One of the main challenges facing health services providers in relation to the COVID-19 crisis is diagnosing and predicting how the disease will develop [9]. To address this, the company partnered with relevant hospital stakeholders and collaborators from a computer science department at a top Nordic university to develop the idea of using imaging to diagnose and predict disease development such as the need for intubation, admission to intensive care unit, or prediction of death, and thereby assisting in resource planning at hospitals. During the initial phases of product development, it became clear to the company that LungX could be based on X-ray scans as they are generally faster, more mobile, and more accessible than CT scans. The fact that many hospitals have mobile X-ray scanners was a big advantage; this meant that healthcare providers could bring the scanner to patients instead of vice versa, making it easier to avoid spreading infection. LungX was therefore built to take X-rays as input and use advanced AI and image analysis to provide real-time information that can assist in resource planning. The input LungX provides to the resource planning process comes in the form of a severity score for COVID-19 patients, a modified Brixia score [79] that quantifies how many of the patient's 12 lung zones are affected by lung edema/consolidation. Moreover, LungX is able to detect six different lung abnormalities (atelectasis, fracture, lesion, pleural effusion, pneumothorax, edema/consolidation). The severity score is calculated when edema/consolidation is detected, and it indicates how widespread the finding is throughout the lungs. The detection of the six different lung abnormalities provides trained medical professionals with complementary information in the evaluation of chest X-rays and is assistive in nature. Besides the detection of lung abnormalities and the severity score, LungX also provides a planning element based on triaging patients. The triage assessment for each incoming patient measures the severity of the pathologies detected and is based on a ranking of the lung abnormalities and the severity score to help support planning of patient care. It is meant to give a quick status overview of current patients and prioritize those with the most severe symptoms. Thus, LungX has multiple value propositions that can be utilized based on the specific needs of a given hospital. For the interested reader, the technical details, data foundation and test performance metrics for LungX are described in Appendix 1.

The information provided by LungX is visualized in two different interfaces: The Study List and the Findings Viewer. The Study List provides an overview of all patients scanned and their triage category. This means that the Study List can be used to prioritize patients based on their triage category so the patients with the most critical triage category can be seen first. The Findings Viewer shows all the information extracted by LungX for a single patient, including their scan, their triage category, and a findings overview showing whether each of the six abnormalities has been found in a binary fashion (present or not present). If a given abnormality has been, it is possible to use a 'Findings Selected' option to see a markup of where in the X-ray LungX has identified the finding. Figures 3 and 4 show a representation of the Study List and Findings Viewer, respectively. The two figures are not identical representations of the actual system, but they accurately portray the type and format of information presented in the GUI.

| Patient Name | Patient ID | Study Date | Modality | Scanner ID | Study Summary | Triage |
|---|---|---|---|---|---|---|
| Anon 1 | 1111 | 11/11/11 | CR | 12121212 | Indication of lesion | Medium 🟡 |
| Anon 2 | 2222 | 11/11/11 | CR | 12121212 | Indication of pleural fusion, atelectasis, edema | High 🔴 |
| Anon 3 | 3333 | 11/11/11 | CR | 12121212 | - | ⚪ |
| Anon 4 | 4444 | 11/11/11 | CR | 12121212 | Indication of lesion | Medium 🟡 |
| Anon 5 | 5555 | 11/11/11 | CR | 12121212 | - | Medium 🟡 |

*Figure 3 Study List*

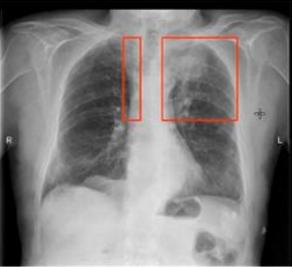

*Figure 4 Findings Viewer*

## 1.5 Findings

While investigating how the need for xAI emerged during the development of the AI application, we observed a plethora of concerns as expressed by the interviewees. In particular, we found clear patterns of variations in how the need for xAI emerged, and we identified four aggregated concerns related to development, domain expertise, decision-making, and patient-related concerns. These concerns manifested with specific stakeholders in mind. In the next sections, we present the specific stakeholder concerns that lead to the need for xAI in AI-based healthcare applications.

### 1.5.1 Development Team

Explanation needs were expressed not only by the members writing the software, building infrastructure, or training the model, but also by people working with and annotating the training data, developing the GUI, and testing the model and product. In this particular setup—the partnership among the university, the company, and the hospital—the company is responsible for researching and developing the best possible solution for identifying and classifying COVID-19 based on CXR (chest X-rays) and potentially other data sources. As a result, we observed that explanation needs emerged through the development team's concerns, which primarily involved Model performance and Data foundation concerns.

*Model Performance Concerns*

Many times, interviewees stated that in order to achieve the aim of the project—to improve the patient journey—the model had to fulfill certain performance requirements while still complying with GDPR policies and local health regulations. The performance of the model constitutes a major development obstacle in terms of obtaining satisfactory accuracy and precision scores.

*"The main obstacle would be to ensure enough, you know, accuracy and precision."* (Medical annotator)

Developers used the explainable element in the GUI to test and validate the model's performance, viewing the output and performance in the GUI to ensure correct visualization of the pathologies in the CXR. Various tests on 20% test data landed around .8- .9 AUC (area under the curve) for detecting COVID-19, green-lighting the model for empirical testing to see if those numbers would still stand. However, two types of pathologies proved more difficult in terms of maintaining a reliable performance. This resulted in a scope creep to four instead of six pathologies in the test version for the clinics and hospitals.

*"… It's not because we don't have enough data, it's just we don't have enough phenotype of the dataset in the sense that you could have lots of pneumothorax pretty clear, you could have faint pneumothorax, but there are some pneumothorax that are extremely hard to see…"* (CTO)

Moreover, the current model only examines images and does not take into account any additional information, such as gender, age or other patient information. On the one hand, this is good in terms of complexity as fewer variables will often be easier to interpret, however, on the other hand the model still needs to maintain a satisfactory performance. This is often a dilemma in developing models for critical decision-making, as one of the interviewers reports that more data would also improve the accuracy of the model:

*"...Right now, you only have the images and do not take into account any other additional patient-related data like age, gender, blood pressure, or whatever it might be that could be relevant in terms of improving the accuracy of the model."* (Developer)

To improve the accuracy of the model, the developers are aware that additional information can significantly influence the model performance.

*"…We did see, well, an increased accuracy when they had age information as well, so I think age was definitely something to include."* (Developer)

However, the current performance proves good enough for testing in empirical settings, but it will be expected to include more vital patient information in testing. Hence, this increases the need for explainability, as complexity in the model increases in parallel with more data and variables to comprehend. Though it's worth noting that LungX is an aiding tool rather than a diagnostics tool, clinicians still need to have some understanding of the model output. However, the use of the model output can differ, depending on the role of the clinician.

*"…For example, a radiologist probably doesn't need as much help finding a series of kind of common pathologies in a chest X-ray, I think they're able to do that really really quickly, versus a junior clinician maybe working in the emergency department who might not have very much experience in looking at X-rays. And that actually might be the main benefit for them, whereas maybe in the radiology department it's not so much about finding the pathology, it's about having something that very quickly scores that and provides a risk assessment."* (Product Owner)

Performance is now a situation-specific measure that depends on the intended use case (e.g., diagnostic assistance or risk assessment) and shows how it can differ from use case to use case. In this sense, performance is not necessarily usefully displayed in numeric values dealing with quick risk assessments, as it is arguably harder to comprehend a numeric value than simple color-coded indications in stressful situations.

Development of the shown output should reflect the stakeholders' needs in their specific situation, including the developers themselves.

*Data foundation concerns*

Data foundation concerns relate to the data collection approach practices of storing and cleaning data, as well as how data is structured, annotated, and evaluated. The ability to build any form of supervised DNN (deep neural network) demands a data foundation that is labelled correctly, with a variation that is evenly distributed. Just as critical is the size of the dataset. Both the variation, size and distribution of the data foundation can be explored with different xAI frameworks to identify potential biases in the data prior to modelling.

It is imperative to have a data set large enough for both training and testing to provide a convincing consistency in the model. In this case, data annotation practices were standardized by an employee with a medical background, who created a protocol for how to interpret the CXR, segment them into different pathologies, and annotate them. Three annotated datasets were used — one received from the hospital (containing COVID-19 CXR) and two open-source datasets containing different types of similar pathologies. These datasets established the data foundation of the model. The severity score the company developed builds upon the brixia scoring system for COVID-19 [79], [80] to segment the lungs into 12 different zones, visualized and explained in the user manual.

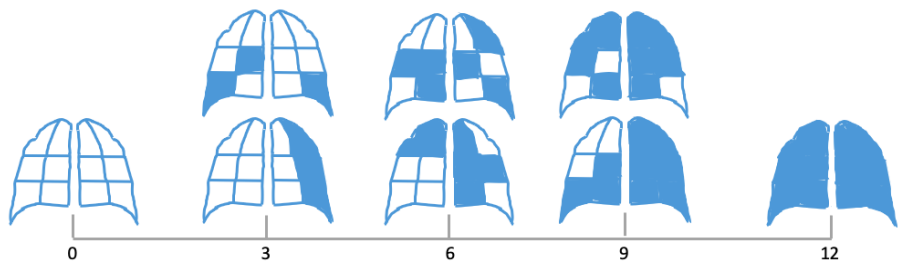

*Figure 5 Visualization of ZASS Calculation*

The zonal assessment severity score (ZASS) and the brixia scores have been used at clinics as ICU doctors need something easily interpretable to measure patient progress. In this case, the company found out that radiologists were doing this by hand. They may, therefore, be familiar with interpreting this scoring, which is now automated for them. As illustrated in Figure 5, the model measures the affected areas of the lungs and gives a total score of 0-12.

*"…And then we found out, at one hospital, the radiologists were scoring the X-rays just because the intensive care doctors wanted something quantifiable to base patient progress on. That helped us define if we should use a similar scoring system. We can automate that scoring."* (CTO)

Moreover, the company has employed several doctors—one of whom is the product owner—to guide the team as to which information is important for ICU doctors and radiologists when they make their decisions. The team has also established an international network of specialists from around the world, including Italy, which was one of the first countries to shut down due to the pandemic. The insights and learnings gained from these partnership-discussions with other clinicians were used to prioritize information and metadata related to the images. Furthermore, these discussions prompted the team to replace the previous heatmap overlays on the X-rays with the current boundary boxes to indicate findings in an image.

However, knowing what data and information you want is not always the same as having it; one of the developers also expressed concern with using open-sourced data, as some information such as which scanner was used, image quality, and if any preprocessing of the images has been done is often not described in these datasets. Besides these concerns, the ability to generate correctly classified COVID-19 instances is another hurdle the team has managed to tackle with an accuracy around .8, by training and educating their employees in reading and interpreting CXR.

### 1.5.2 Subject Matter Expert

Subject matter expert (SME) concerns relate to domain knowledge and validation practices. The role of the SME can shift depending on the context. In the development phase of LungX, when annotating the images with the different pathologies, the SMEs consist of the employed clinicians and one radiologist. The annotation team was creating the ground truth for the machine learning model to be able to recognize COVID-19. As such, the team has gone through extensive training in detecting abnormalities in CXR. Even though some are doctors themselves, the team members are young and have therefore chosen to partner up with a more senior radiologist with years of experience

*Validation Concerns*

The team is very cautious of the life-or-death stakes their product is dealing with, which reflects their validation concerns and how the company addresses the need for a sound model that supports the trust of all stakeholders. Operating in high-stakes environments has led the team to maintain a human-in-the-loop approach when inspecting the annotations of the data foundation, even though machine learning helped in identifying many of the pathologies by using NLP (natural language processing) tools to identify pathologies of interest. It's worth noting that the more uncertainties the model relies on, the more doubtful it will be in the end, though this will not necessarily be visible in the performance metrics. To approach this, the company provided manual insurance by having SMEs oversee the annotation process.

*"…When we look at the images, we also had a report from the radiologists who had described it previously, and we use that report as a ground truth for confirming that findings were there. And the way that those images got found was an algorithm search in these reports. And they found keywords such as pneumothorax or pneumonia. And then they took that image and put it into the pneumonia, or pneumothorax dataset."* (Medical annotator)

Besides the validation of training data, the company faces a clinical validation process based on images delivered by the hospital. In this setting, the SME would be one of the collaborators from the hospital reviewing the output of the model and evaluating if the results are substantial. If so, they will continue testing the product in production.

*"... There will be a process of formal clinical validation happening in the next couple of weeks, and so that will be an analysis based on a data set and dedicated data set for validation..."* (Medical annotator)

*"... The clinical validation part is together with the clinics finding out whether or not the LungX solution is relevant for [the clinics] and actually provides benefit for them...and obviously, Hospital X is more a partner in terms of providing medical insights into what is needed and necessary, but also for testing and validation."* (CEO)

Validating the output of the model with experienced professionals from the hospital as well as medical annotators led the company to drop two of the original six pathologies in scope, as their performance was not living up to their standards.

### 1.5.3 Decision-Makers

Decision-making concerns are very prominent among the interviewees, as the value proposition of the product is directly related to the use of the system for improving decisions in the hospitals. However, LungX entails not only one but multiple value propositions supporting different end users. These could be the ICU clinicians, the hospital manager, the radiologist, or the nurse. For instance, the severity score (ZASS) would primarily be useful for the ICU clinicians. We identified four themes which represent concerns about decision-makers (e.g. clinicians, radiologist) and their interaction with LungX. These concerns are inferred from the interviews with other stakeholders both medical and development staff at the company.

***Human/AI Discretionary Boundaries***

Decision-Making is a process where various information is collected or presented, interpreted, and used by humans to draw a conclusion. In the case of using AI systems to assist this process, the interviewees express their concern regarding discretionary boundaries between the model output and the decision-maker's judgement.

*"I'm always curious about what [LungX] says. I have my own findings, and I'm also very curious about what it says. If it's the same, then I'm more confident in my findings. So as a young doctor, I would use it for that. Even though it can't catch it all."* (Medical annotator)

*"Trying to understand who's liable for the decisions is extremely important."* (CEO)

The user interface of LungX has limited information on how certain the model is in its classification, which could lead young doctors to rely too much on potentially low accuracy output from the model. The need to adjust the complexity of the information displayed could be re-evaluated; though there is a risk of contaminating the interface, an accuracy score would potentially awaken a more critical judgement in young doctors. Hence, the need for further understanding the interplay of a human and AI in order to achieve altogether better diagnostic outcomes. This information is consistently requested by the employees and is on the roadmap for implementation.

*"I like to know the boundaries, the thresholds that determine whether the machine learning model classified it as normal or not normal. I'd like to know that threshold so I can use that in my own decision-making. And then I regard the software, the*

*framework, as helping me with some inputs to my own decisions. That's why I don't want it to [automatically] make decisions for me, like diagnosing."* (Medical annotator)

Though LungX will influence the decision-making process, it is emphasized several times that none of the clinical staff at the company wish for the model to take over the decision; rather, they'd prefer to use it as an aiding tool in their own decision-making. Therefore, it is even more important to stress that the final decision belongs to the doctor and that the model cannot stand alone.

*"The primary goal is to assist with diagnostics or assist with planning but not kind of take over those processes because that's not the goal, and I think there's much more risk when software tries to automate processes in medicine, and that's not what we're trying to do."* (CEO)

Moreover, the tool does not take into account the patient's history or any information from the clinical investigation (blood samples, blood pressure, etc.), all of which are highly relevant to the final decision.

*"I deem that the clinical information, the patient history, and the clinical signs and the results of the clinical investigation of the patient is very important in telling me what it might be that I'm looking at, and so that's important to include. It wasn't included in this product and it isn't included right now, but I think that's actually really important in doing that."* (Medical annotator)

For the decision-maker, the complexity lies in judging when and how they should factor the information presented in LungX into their own decision-making. Depending on where these boundaries are agreed upon, different xAI needs emerge. The more complex and autonomous the system, the more need for information that explains the 'why.' Furthermore, the tradeoff between complexity and interpretability emerges as the decision-making workflow should be smooth and undisturbed while still generating an informative, reliable explanation.

### *Output Interpretation Concerns*

Whereas the theme of discretionary boundaries is concerned with when and what the decision-maker should use the aiding tool in their decision-making process, the theme output interpretation is about how the information is presented and interpreted by them. The information presented in the interface is intended to be easily interpreted, stripped of unnecessary noise to support a smooth and fast process. The most prominent feature in the software is the red 'bounding boxes' that enclose abnormalities in the image, their findings are then defined in a table beside the image as shown in the illustration of the Findings Viewer.

*"It's basically some sort of explanation of what we think [bounding box]. If there's a 95% probability of seeing a pleural effusion, the radiologist will think OK, why? Therefore, we have to be able to show exactly the reason and the region in which we think there is a high probability of a pleural effusion. So, this is where the bounding boxes come into the picture, to show the radiologist that these are the regions we think are the most suspicious of a particular finding."* (CTO)

Though the probability score of the findings (i.e., how certain the model is of its classification) is not displayed in the interface, information on the severity of the finding is shown (i.e., how widespread the disease detected in the lungs is). The severity score is considered more difficult to comprehend, as it consists of several underlying factors, such as the severity ranking of the detectable diseases—as decided by the individual clinic—and the ZASS score, which is based on the brixia scoring system for COVID-19 to measure the extent of the disease [79]. As the severity score is subject to adjustment from clinic to clinic depending on their practices, this could increase the complexity in interpretation of the score.

*"I think that that it [bounding box] mitigates a lot of that misinterpretation risk. At the moment, I think it's relatively easy to interpret. You know, if there's a finding or if there's not, but in terms of actually… the fact that it actually gives a score to quantify how widespread a finding is. I mean, that then can be misinterpreted, and so we have, for example, an explanation of what that score means in the user manual…I think it is hard to communicate risk, particularly when that would then involve some clinical information as well, and then that becomes less easy to explain to a user because you can't just say, 'Oh, the lung was divided into the zones and, therefore, we kind of get this score."* (Product Owner)

How the clinicians interpret the visual information displayed is considered relatively easy when discussing the images and their boundary boxes. Moreover, the red/yellow/green triage field showing which patient should be prioritized first according to severity of disease and type of disease is deemed understandable for prioritizing patients but not for supporting diagnosis.

*"In our software, we also have a triage function, and that's very much low, medium, high triage. And I think that that is a pretty simple way to communicate risk or urgency to a user."* (Product Owner)

The quantified severity scores are deemed somewhat more complicated. Because of this, the company has incorporated two measures to accommodate that: training sessions for users on using and interpreting information in LungX to ensure that decision-makers feel comfortable using the values displayed.

*"I think also there are some risks related to more subtle things like interpretation of results […] what do the results mean, and I think a lot of that has to then come down to making sure that everyone who uses the technology is trained in how to use it."* (Product Owner)

Moreover, the current decision-making process involves interaction between colleagues if they doubt a claim made by another clinician. This is seen as difficult to maintain when operating with AI-based systems.

*"I could ask my colleague, 'Why do you think that this person has pneumonia?' And he will answer [X, Y, Z], and OK, then I'll trust him if I agree, but I can't really ask the network [LungX]."* (Medical annotator)

### *Adoption Concerns*

However, it is not enough to know how and what to use in a decision-making process if the product is not adopted in the decision-making process. The adoption concerns theme refers to how the employees experience resistance toward the product or express doubt about how it will be received and used.

*"Purely, it comes down to the lack of being able to get a good way of applying it in the clinic. That is the last step that fails because you might be able to sort of say, 'OK, there's a very good AI, it works really well, it has high accuracy, good sensitivity, specificity.'"* (Clinical Operations Officer)

According to interviewees, understanding the processes and workflow surrounding the work of an AI-based model is a key concern. The way LungX is incorporated into the overall process influences its adoption in hospitals. Interviewees also express a more general concern with using automation tools within medicine, as this field often deals with high-stake decisions.

*"So, I think it's twofold. I think it's that you need to make sure that performance is very, very good. But I also think… it's just very difficult for people to accept automation within medicine I think."* (Clinical Operations Officer)

Though LungX is not intended for automation, it has a significant influence on the decision-making process. The development team seems to be aware of potential resistance from medical staff, especially when introduced to new and unknown tools. LungX is one of the first systems of its kind to be implemented in clinics, which may subject it to resistance in many forms.

In addition, some clinics might operate with contingencies from other software products that inhibit them to change or alter their current processes. Moreover, the LungX product might not fit exactly what the clinics are looking for. Different pathologies can be of interest to different clinics, so the team risks having spent time on developing a product that is not relevant in real-life scenarios.

*"It's able to differentiate A from B. You go to the clinic and they say, 'Well, we can't implement that because, you know, we use this different product and we can't change that because the whole region has bought this product. We don't care about A or B. We care whether it's A&B or C.' So, you solve the wrong problem or maybe it's the wrong people that have to implement it. That is where you finally realize you've got a hammer and there's no nails that match it."* (Clinical Operations Officer)

This proves that deep knowledge of how the product is intended for use in a real-life scenario can be essential in order to ensure adoption by stakeholders in general. Moreover, the information required by the decision-makers needs to be present for them to benefit from the tool, wherefore the company has initiated the collaboration with hospitals and clinics worldwide. Here, they have acquired knowledge on what kind of information is relevant to stakeholders. However, a wide range of factors come into play in the adoption of new tools, including trust in the product, culture, local work processes, and general support from management.

*Trust Concerns*

The decision-maker has to make the final call about a patient's health and further treatment; understandably, this leads them to be very cautious in their decision-making, including the information they rely on to make their decision. After all, they are held liable and face the highest risks, both personally and professionally. Because using AI in decision-making within medical imaging is still a fairly new concept, the trust factor remains fragile, as with all new things.

*"… To have other doctors who are less enthusiastic about new technology, to have them use it because if the software, for instance, makes one mistake, just one mistake, it could be fatal, then the trust in it will be almost lost."* (Medical annotator)

To support public trust in the model for those who need more evidence of reliability, the company has scheduled clinical trials. Moreover, the implementation of xAI in the form of the 'bounding boxes' and the color coding for the triage score, makes the interface amenable to decision-makers such as radiologists or ICU doctors.

*"I think people are quite nervous around this technology, and I think you kind of need the experiences from those trials in hospitals to convince people to give things a go."* (CEO)

*"And the clinical evaluation reports, the clinical trials that we are going to run with the hospitals that will establish the reliability of the system."* (CTO)

Though LungX is not automating the entire diagnostic process, it provides information that must be trusted in order to be useful. Interviewees explain that they discuss different cases when in doubt, as they are never always right. The same goes for AI models, wherefore it should never be blindly trusted. Some doctors may rely too much on the information provided and end up letting the model reinforce potentially flawed decisions.

*"...Maybe the less experienced doctors would trust it so much that they wouldn't have a second opinion on the image."* (Medical annotator)

*"If the radiologist blindly trusts the algorithm without even looking at the images and just ships the image of... said OK, there's a lesion in there, then the 'bug' is on the radiologist's desk, and this is why understanding liability is so important — who's liable for it?"* (CTO)

All interviewees stressed the fact that LungX is an aiding tool and cannot be held liable in any way, making the radiologist describing the images and doing the diagnostics report the liable party. The risk of being liable, as a software provider, for an undetected case of COVID-19 could have fatal consequences. Hence, accountability boundaries need to be made explicit that would allow the decision maker to critically use the output of the ML by applying their clinical judgment.

### 1.5.4 Audience

Since the case study investigates the development of the LungX product, the typical audience discussed in the interviews was patients themselves. In the end, clinicians and radiologists are trying to treat and save patients' lives, and LungX is intended to improve the patient journey by assisting in speedy identification of pneumonia related COVID-19 in the lungs and its severeness. The interviewees expressed concerns related to mitigating decision consequences for the patients.

***Decision Consequences***

The value proposition was to use LungX to quickly identify which patients need highest priority and which patients needed to be admitted to the hospital. LungX is dealing with high-stakes decisions that involve life or death for patients. In this sense, the interviewees are aware of the severity and how it can improve *"...the patient journey in terms of disease progression and predict the need for Intensive Care Unit, hospitalization, ICU or even ventilation, respiratory support or death"* (CEO)

All interviewees express a high level of concern for the patient and managing the high stakes, both in terms of prioritizing the most severe cases and determining how individual cases should be dealt with according to patient history. The high stakes are reflected throughout the model development in the product's great attention to detail, high-level performance, and meticulous process for validating the model's training data.

*"...At the end of the day, it is a tool that interferes with a patient's life, so to speak... You can't call back a patient without having consequences, so this is really important to understand... If you have a patient that's diagnosed wrongly, and the patient gets some sort of an attack, then it's a lifelong rehabilitation."* (CTO)

Presenting LungX as a decision support tool where the decision-maker is still the liable party—rather than a completely automated diagnostic tool—ensures that the patient will preserve the possibility to object to a decision or ask for an explanation. In this case, the company guards this possibility due to the severity of the matter.

## 1.6 Discussion and Future Work

In this study, we investigated the development efforts of LungX, an AI-based model for diagnosing COVID-19 in patients. In this context, we identified a plethora of xAI needs that emerge during the idea formulation and development phases. Further, we found that these xAI needs emerged through four aggregated concerns related to development, domain expertise, decision-making, and audience-related concerns. We also found that all these concerns were expressed either by a specific stakeholder or with a specific stakeholder group in mind (see figure 6).

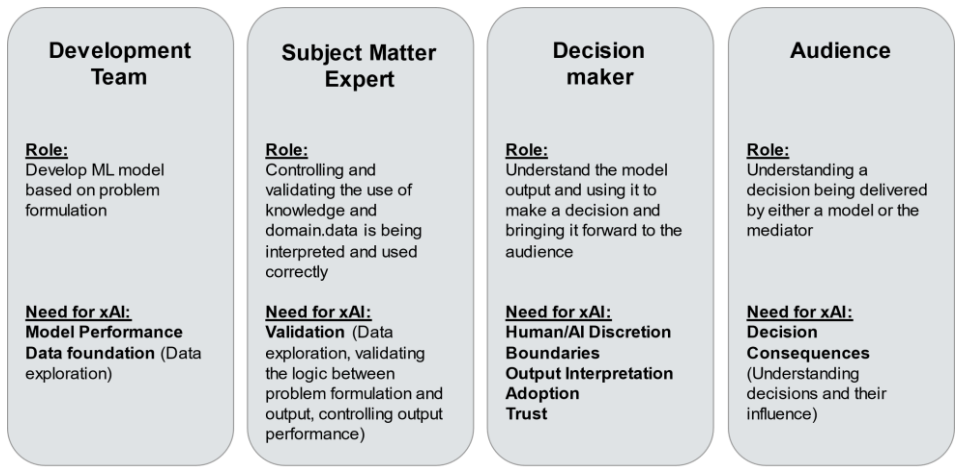

*Figure 6 Concerns that xAI can alleviate depend on the stakeholder group*

Explanation concerns expressed by and for the development team have been highly recognized in the literature [6], [47], [81] where developers use xAI frameworks to make sense of the datasets used as well as the inner workings of AI-based models in order to achieve performance results. Similarly, our findings in relation to xAI emerging through decision-making concerns—particularly manifesting in output interpretation as well as adoption and trust concerns—support earlier studies that found ambiguity as an effect of using AI-based healthcare applications [11] or resistance to using these applications [2]. Further, it becomes apparent that the need for xAI also emerges from trying to establish discretionary boundaries, a point linked to the problem of accountability when using AI-based systems [82]. xAI needs driven by subject matter expert concerns refer to the necessity of domain knowledge during the development of AI applications [83]. Finally, audience related concerns drive xAI needs in order to be able to communicate decisions to patients as well as manage and mitigate decision consequences that may severely impact patients [84].

By highlighting the different concerns as represented by different stakeholder groups, our study provides empirical evidence of multiple stakeholder xAI needs and concerns in relation to AI — in contrast to the usual developer-user perspective [24]. Building more visual explanations may resolve some of these concerns; however, some stakeholder groups might not be satisfied with a visualization of a finding in a picture, or red/yellow/green dots indicating the ideal prioritization of patients. Instead, they may require more scientific explanations entailing more complexity [85]. Our study shows that the tradeoff between explanation accuracy and comprehensibility varies among different stakeholders. It appears that the closer a stakeholder works with the AI model, the more accurate an explanation they need. This is not a problem for developers, since their knowledge on AI systems, statistics, and computer science is extensive. Our findings indicate that during the development of LungX, the tradeoff point is carefully considered for subject matter experts and decision-makers (radiologists and ICU clinicians). Subject matter experts and decision-makers need both accurate and comprehensible explanations to be able to validate an AI system; these stakeholders oftentimes lack basic knowledge of AI systems, yet they are expected to be able to validate and rely on them in order to do their work. On one hand, they need accurate explanations to make confident decisions based on the AI output. On the other hand, they also need comprehensible explanations both to trust the system themselves as well as to be able to communicate their decision to the patient. Hence, they need models to offer comprehensible explanations that do not sacrifice accuracy. Finally, for the audience, the balance seems to shift toward more comprehensible explanations that may sacrifice some degree of thorough accuracy. As such, our study responds to recent calls for a stakeholder perspective in xAI research [24], [46].

The plethora of concerns from each stakeholder group further reinforces the need for a multidisciplinary approach when developing xAI solutions, as recent conceptual works have called for [24], [26], [35], [36]. Moreover, the study shows a continued demand for understanding the workflow around the different stakeholders to address their specific explanation needs due to output interpretation concerns. The model might be right, but results interpreted and used incorrectly can result in biased decisions and unintended treatment, as in the case of COMPAS.

Our findings, however, focus on the emergence of xAI needs during the development of AI-based systems. Future studies should also investigate xAI during the implementation and use of these AI-based applications to address emerging xAI needs, especially as existing and additional stakeholders become more active and have more in-situ insights.

While our findings emphasized how the need for xAI emerged during development, future studies could advance our knowledge of how different technical frameworks can assist these stakeholder concerns.

**References**

[1] A. B. Arrieta *et al.*, "Explainable Artificial Intelligence (XAI): Concepts, Taxonomies, Opportunities and Challenges toward Responsible AI," *Inf. Fusion*, no. October, 2019.

[2] C. J. Cai, S. Winter, D. Steiner, L. Wilcox, and M. Terry, "'Hello Ai': Uncovering the onboarding needs of medical practitioners for human–AI collaborative decision-making," *Proceedings of the ACM on Human-Computer Interaction*, vol. 3, no. CSCW. Association for Computing Machinery, pp. 1–24, 01-Nov-2019.

[3] A. Adadi and M. Berrada, "Peeking Inside the Black-Box: A Survey on Explainable Artificial Intelligence (XAI)," *IEEE Access*, vol. 6, pp. 52138–52160, 2018.

[4] F. Doshi-velez and B. Kim, "Towards A Rigorous Science of Interpretable Machine Learning," no. Ml, pp. 1–13, 2017.

[5] Z. C. Lipton, "The Mythos of Model Interpretability," no. Whi, Jun. 2016.

[6] M. T. Ribeiro, S. Singh, and C. Guestrin, "'Why should i trust you?' Explaining the predictions of any classifier," in *Proceedings of the ACM SIGKDD International Conference on Knowledge Discovery and Data Mining*, 2016, vol. 13-17-Augu, pp. 1135–1144.

[7] A. Holzinger, C. Biemann, C. S. Pattichis, and D. B. Kell, "What do we need to build explainable AI systems for the medical domain?," no. Ml, pp. 1–28, 2017.

[8] U. Pawar, D. O'Shea, S. Rea, and R. O'Reilly, "Incorporating explainable artificial intelligence (XAI) to aid the understanding of machine learning in the healthcare domain," *CEUR Workshop Proc.*, vol. 2771, no. December, pp. 169–180, 2020.

[9] J. Phua *et al.*, "Intensive care management of coronavirus disease 2019 (COVID-19): challenges and recommendations," *The Lancet Respiratory Medicine*, vol. 8, no. 5. Lancet Publishing Group, pp. 506–517, 01-May-2020.

[10] E. Tjoa and C. Guan, "A Survey on Explainable Artificial Intelligence (XAI): Towards Medical XAI," vol. 1, 2019.

[11] S. Lebovitz, "Diagnostic doubt and artificial intelligence: An inductive field study of radiology work," *40th Int. Conf. Inf. Syst. ICIS 2019*, 2020.

[12] T. Davenport and R. Kalakota, "The potential for artificial intelligence in healthcare," *Futur. Healthc. J.*, vol. 6, no. 2, pp. 94–98, 2019.

[13] T. Panch, H. Mattie, and L. A. Celi, "The 'inconvenient truth' about AI in healthcare," *npj Digit. Med.*, vol. 2, no. 1, pp. 4–6, 2019.

[14] A. L. Fogel and J. C. Kvedar, "Artificial intelligence powers digital medicine,"

*npj Digit. Med.*, vol. 1, no. 1, pp. 3–6, 2018.

[15] W. Hryniewska, P. Bombiński, P. Szatkowski, P. Tomaszewska, A. Przelaskowski, and P. Biecek, "Do not repeat these mistakes -- a critical appraisal of applications of explainable artificial intelligence for image based COVID-19 detection," no. January, 2020.

[16] C. O'Neil, *Weapons of Math Destruction: How Big Data Increases Inequality and Threatens Democracy*, vol. 272. 2016.

[17] A. Shaban-Nejad, M. Michalowski, and D. L. Buckeridge, "Health intelligence: how artificial intelligence transforms population and personalized health," *npj Digit. Med.*, vol. 1, no. 1, 2018.

[18] A. Sharma, S. Rani, and D. Gupta, "Artificial Intelligence-Based Classification of Chest X-Ray Images into COVID-19 and Other Infectious Diseases," *Int. J. Biomed. Imaging*, vol. 2020, 2020.

[19] E. Strickland, "How IBM Watson Overpromised and Underdelivered on AI Health Care - IEEE Spectrum," 2019. [Online]. Available: https://spectrum.ieee.org/biomedical/diagnostics/how-ibm-watson-overpromised-and-underdelivered-on-ai-health-care. [Accessed: 26-Jan-2021].

[20] Z. Obermeyer, B. Powers, C. Vogeli, and S. Mullainathan, "Dissecting racial bias in an algorithm used to manage the health of populations," *Science (80-. ).*, vol. 366, no. 6464, pp. 447–453, Oct. 2019.

[21] W. D. Heaven, "Google's medical AI was super accurate in a lab. Real life was a different story. | MIT Technology Review." [Online]. Available: https://www.technologyreview.com/2020/04/27/1000658/google-medical-ai-accurate-lab-real-life-clinic-covid-diabetes-retina-disease/. [Accessed: 16-Mar-2021].

[22] J. Angwin, J. Larson, S. Mattu, and L. Kirchner, "Machine Bias — ProPublica," *ProPublica*. [Online]. Available: https://www.propublica.org/article/machine-bias-risk-assessments-in-criminal-sentencing. [Accessed: 03-Mar-2019].

[23] T. W. Kim, "Explainable artificial intelligence (XAI), the goodness criteria and the grasp-ability test," pp. 1–7, 2018.

[24] J. Gerlings, A. Shollo, and I. D. Constantiou, "Reviewing the Need for Explainable Artificial Intelligence (xAI)," in *HICSS 54*, 2021, pp. 1284–1293.

[25] J. Kemper and D. Kolkman, "Transparent to whom? No algorithmic accountability without a critical audience," *Inf. Commun. Soc.*, 2019.

[26] D. D. Miller, "The medical AI insurgency: what physicians must know about data to practice with intelligent machines," *npj Digit. Med.*, vol. 2, no. 1, Dec. 2019.

[27] J. Burrell, "How the machine 'thinks': Understanding opacity in machine learning algorithms," *Big data Soc.*, vol. January-Ju, pp. 1–12, 2016.

[28] A. Páez, "The Pragmatic Turn in Explainable Artificial Intelligence (XAI)," *Minds Mach.*, 2019.

[29] M. BHANDARI and D. JASWAL, "Decision Making in Medicine-An Algorithmic Approach," *Med. J. Armed Forces India*, 2002.

[30] J.-B. Lamy, B. Sekar, G. Guezennec, J. Bouaud, and B. Séroussi, "Explainable

artificial intelligence for breast cancer: A visual case-based reasoning approach," *Artif. Intell. Med.*, vol. 94, pp. 42–53, Mar. 2019.

[31] A. Wodecki *et al.*, "Explainable Artificial Intelligence ( XAI ) The Need for Explainable AI," *Philos. Trans. A. Math. Phys. Eng. Sci.*, 2017.

[32] S. M. Lauritsen *et al.*, "Explainable artificial intelligence model to predict acute critical illness from electronic health records," 2019.

[33] N. Prentzas, A. Nicolaides, E. Kyriacou, A. Kakas, and C. Pattichis, "Integrating machine learning with symbolic reasoning to build an explainable ai model for stroke prediction," in *Proceedings - 2019 IEEE 19th International Conference on Bioinformatics and Bioengineering, BIBE 2019*, 2019.

[34] S. Lebovitz, H. Lifshitz-Assaf, and N. Levina, "To Incorporate or Not to Incorporate AI for Critical Judgments: The Importance of Ambiguity in Professionals' Judgment Process." 15-Jan-2020.

[35] A. Rajkomar, J. Dean, and I. Kohane, "Machine Learning in Medicine," *N. Engl. J. Med.*, vol. 380, no. 14, pp. 1347–1358, 2019.

[36] M. T. Keane and E. M. Kenny, "How Case-Based Reasoning Explains Neural Networks: A Theoretical Analysis of XAI Using Post-Hoc Explanation-by-Example from a Survey of ANN-CBR Twin-Systems," in *Lecture Notes in Computer Science (including subseries Lecture Notes in Artificial Intelligence and Lecture Notes in Bioinformatics)*, 2019.

[37] Stanford University, "Artificial Intelligence and Life in 2030," p. 52, 2016.

[38] L. Reis, C. Maier, J. Mattke, M. Creutzenberg, and T. Weitzel, "Addressing User Resistance Would Have Prevented a Healthcare AI Project Failure," *MIS Q. Exec.*, vol. 19, no. 4, pp. 279–296, 2020.

[39] A. Vellido, "The importance of interpretability and visualization in machine learning for applications in medicine and health care," *Neural Comput. Appl.*, vol. 32, no. 24, pp. 18069–18083, Dec. 2020.

[40] A. Holzinger, G. Langs, H. Denk, K. Zatloukal, and H. Müller, "Causability and explainability of artificial intelligence in medicine," *Wiley Interdisciplinary Reviews: Data Mining and Knowledge Discovery*, vol. 9, no. 4. pp. 1–13, 2019.

[41] M. Goldstein and S. Uchida, "A comparative evaluation of unsupervised anomaly detection algorithms for multivariate data," *PLoS One*, vol. 11, no. 4, pp. 1–31, 2016.

[42] C. Molnar, "Interpretable Machine Learning. A Guide for Making Black Box Models Explainable.," *Book*, p. 247, 2019.

[43] Z. C. Lipton, "The Mythos of Model Interpretability," *Commun. ACM*, vol. 61, no. Whi, pp. 35–43, Jun. 2016.

[44] G. Ciatto, M. I. Schumacher, A. Omicini, and D. Calvaresi, "Agent-Based Explanations in AI: Towards an Abstract Framework," in *Lecture Notes in Computer Science (including subseries Lecture Notes in Artificial Intelligence and Lecture Notes in Bioinformatics)*, 2020, vol. 12175 LNAI, pp. 3–20.

[45] L. H. Gilpin, D. Bau, B. Z. Yuan, A. Bajwa, M. Specter, and L. Kagal, "Explaining explanations: An overview of interpretability of machine learning," *Proc. - 2018 IEEE 5th Int. Conf. Data Sci. Adv. Anal. DSAA 2018*,

pp. 80–89, 2019.

[46] T. Miller, "Explanation in Artificial Intelligence : Insights from the Social Sciences," 2018.

[47] A. Goldstein, A. Kapelner, J. Bleich, and E. Pitkin, "Peeking Inside the Black Box: Visualizing Statistical Learning With Plots of Individual Conditional Expectation," *J. Comput. Graph. Stat.*, vol. 24, no. 1, pp. 44–65, 2015.

[48] R. Brandão, J. Carbonera, C. de Souza, J. Ferreira, B. N. Gonçalves, and C. F. Leitão, "Mediation Challenges and Socio-Technical Gaps for Explainable Deep Learning Applications," *arXiv Prepr. arXiv …*, no. Query date: 2020-04-16 13:43:28, pp. 1–39, 2019.

[49] O. Biran and C. Cotton, "Explanation and Justification in Machine Learning: A Survey," *IJCAI Work. Explain. AI*, no. August, pp. 8–14, 2017.

[50] S. T. Mueller, R. R. Hoffman, W. Clancey, A. Emrey, and G. Klein, "Explanation in Human-AI Systems: A Literature Meta-Review," *Def. Adv. Res. Proj. Agency*, no. February 2019, p. 204, 2019.

[51] R. Guidotti, A. Monreale, S. Ruggieri, F. Turini, F. Giannotti, and D. Pedreschi, "A survey of methods for explaining black box models," *ACM Comput. Surv.*, vol. 51, no. 5, 2018.

[52] A. Asatiani, P. Malo, P. R. Nagbøl, E. Penttinen, T. Rinta-Kahila, and A. Salovaara, "Challenges of Explaining the Behavior of Black-Box AI Systems," *MIS Q. Exec.*, vol. 19, no. 4, pp. 259–278, 2020.

[53] T. Miller, P. Howe, and L. Sonenberg, "Explainable AI: Beware of Inmates Running the Asylum," 1990.

[54] P. Madumal, L. Sonenberg, T. Miller, and F. Vetere, "A grounded interaction protocol for explainable artificial intelligence," in *Proceedings of the International Joint Conference on Autonomous Agents and Multiagent Systems, AAMAS*, 2019, vol. 2, pp. 1033–1041.

[55] W. Samek, T. Wiegand, and K.-R. Müller, "Explainable Artificial Intelligence: Understanding, Visualizing and Interpreting Deep Learning Models," Aug. 2017.

[56] R. R. Hoffman, S. T. Mueller, G. Klein, and J. Litman, "Metrics for Explainable AI: Challenges and Prospects," pp. 1–50, 2018.

[57] Z. Che, S. Purushotham, R. Khemani, and Y. Liu, "Interpretable Deep Models for ICU Outcome Prediction," *AMIA … Annu. Symp. proceedings. AMIA Symp.*, vol. 2016, no. August, pp. 371–380, 2016.

[58] M. Ilyas, H. Rehman, and A. Nait-ali, "Detection of Covid-19 From chest x-ray images using artificial intelligence: an early review," *arXiv*, pp. 1–8, 2020.

[59] C. H. Sudre *et al.*, "Anosmia and other SARS-CoV-2 positive test-associated symptoms, across three national, digital surveillance platforms as the COVID-19 pandemic and response unfolded: An observation study," *medRxiv*. medRxiv, 2020.

[60] E. Dong, H. Du, and L. Gardner, "An interactive web-based dashboard to track COVID-19 in real time," *The Lancet Infectious Diseases*, vol. 20, no. 5. Lancet Publishing Group, pp. 533–534, 01-May-2020.

[61] Y. Hswen, J. S. Brownstein, X. Xu, and E. Yom-Tov, "Early detection of

COVID-19 in China and the USA: Summary of the implementation of a digital decision-support and disease surveillance tool," *BMJ Open*, vol. 10, no. 12, Dec. 2020.

[62] T. Macaulay, "AI sent first coronavirus alert, but underestimated the danger," *The Next Web*, 2020. [Online]. Available: https://thenextweb.com/neural/2020/02/21/ai-sent-first-coronavirus-alert-but-underestimated-the-danger/. [Accessed: 09-Jan-2021].

[63] M. E. H. Chowdhury *et al.*, "Can AI Help in Screening Viral and COVID-19 Pneumonia?," *IEEE Access*, vol. 8, pp. 132665–132676, 2020.

[64] J. Bullock, A. Luccioni, K. Hoffman Pham, C. Sin Nga Lam, and M. Luengo-Oroz, "Mapping the landscape of Artificial Intelligence applications against COVID-19," *J. Artif. Intell. Res.*, vol. 69, pp. 807–845, 2020.

[65] K. Murphy *et al.*, "COVID-19 on chest radiographs: A multireader evaluation of an artificial intelligence system," *Radiology*, vol. 296, no. 3, pp. E166–E172, 2020.

[66] J. Zhang *et al.*, "Viral Pneumonia Screening on Chest X-rays Using Confidence-Aware Anomaly Detection," *IEEE Trans. Med. Imaging*, pp. 1–1, 2020.

[67] X. Li, C. Li, and D. Zhu, "COVID-MobileXpert: On-Device COVID-19 Patient Triage and Follow-up using Chest X-rays," 2020.

[68] J. D. Arias-Londoño, J. A. Gomez-Garcia, L. Moro-Velazquez, and J. I. Godino-Llorente, "Artificial Intelligence applied to chest X-Ray images for the automatic detection of COVID-19. A thoughtful evaluation approach," pp. 1–17, 2020.

[69] R. M. Wehbe *et al.*, "DeepCOVID-XR: An Artificial Intelligence Algorithm to Detect COVID-19 on Chest Radiographs Trained and Tested on a Large US Clinical Dataset," *Radiology*, p. 203511, 2020.

[70] R. R. Selvaraju, M. Cogswell, A. Das, R. Vedantam, D. Parikh, and D. Batra, "Grad-CAM: Visual Explanations from Deep Networks via Gradient-based Localization," *Int. J. Comput. Vis.*, vol. 128, no. 2, pp. 336–359, Oct. 2016.

[71] M. M. Ahsan, K. D. Gupta, M. M. Islam, S. Sen, M. L. Rahman, and M. S. Hossain, "Study of different deep learning approach with explainable AI for screening patients with covid-19 symptoms: Using CT scan and chest X-ray image dataset," *arXiv*, 2020.

[72] K. Conboy, G. Fitzgerald, and L. Mathiassen, "Qualitative methods research in information systems: Motivations, themes, and contributions," *European Journal of Information Systems*, vol. 21, no. 2. Palgrave, pp. 113–118, 24-Mar-2012.

[73] P. Powell and G. Walsham, "Interpreting Information Systems in Organizations.," *J. Oper. Res. Soc.*, 1993.

[74] A. George and A. Bennett, *Case Studies and Theory Development in the Social Science*. Cambridge, MA: MIT Press, 2005.

[75] J. M. Bartunek, P. G. Foster-Fishman, and C. B. Keys, "Using collaborative advocacy to foster intergroup cooperation: A joint insider-outsider investigation," *Hum. Relations*, vol. 49, no. 6, pp. 701–733, 1996.

[76] D. A. Gioia, K. G. Corley, and A. L. Hamilton, "Seeking Qualitative Rigor in Inductive Research: Notes on the Gioia Methodology," *Organ. Res. Methods*, vol. 16, no. 1, pp. 15–31, 2013.

[77] K. G. Corley and D. A. Gioia, "Identity ambiguity and change in the wake of a corporate spin-off," *Administrative Science Quarterly*, vol. 49, no. 2. pp. 173–208, Jun-2004.

[78] J. Corbin and A. Strauss, *Basics of Qualitative Research (3rd ed.): Techniques and Procedures for Developing Grounded Theory*. SAGE Publications, Inc., 2012.

[79] A. Borghesi and R. Maroldi, "COVID-19 outbreak in Italy: experimental chest X-ray scoring system for quantifying and monitoring disease progression," *Radiol. Medica*, vol. 125, no. 5, pp. 509–513, May 2020.

[80] A. Borghesi *et al.*, "Radiographic severity index in COVID-19 pneumonia: relationship to age and sex in 783 Italian patients," *Radiol. Medica*, vol. 125, no. 5, pp. 461–464, May 2020.

[81] S. M. Lundberg and S. I. Lee, "A unified approach to interpreting model predictions," *Adv. Neural Inf. Process. Syst.*, vol. 2017-Decem, no. Section 2, pp. 4766–4775, 2017.

[82] M. Veale, M. Van Kleek, and R. Binns, "Fairness and accountability design needs for algorithmic support in high-stakes public sector decision-making," in *Conference on Human Factors in Computing Systems - Proceedings*, 2018.

[83] M. Kuzba and P. Biecek, "What would you ask the machine learning model? identification of user needs for model explanations based on human-model conversations," *arXiv*. 2020.

[84] E. Rader, K. Cotter, and J. Cho, "Explanations as mechanisms for supporting algorithmic transparency," in *Conference on Human Factors in Computing Systems - Proceedings*, 2018.

[85] S. Chari, D. M. Gruen, O. Seneviratne, and D. L. McGuinness, "Directions for Explainable Knowledge-Enabled Systems," no. March, 2020.

[86] G. Huang, Z. Liu, L. van der Maaten, and K. Q. Weinberger, "Densely Connected Convolutional Networks," *Proc. - 30th IEEE Conf. Comput. Vis. Pattern Recognition, CVPR 2017*, vol. 2017-January, pp. 2261–2269, Aug. 2016.

[87] R. Selvan *et al.*, "Lung Segmentation from Chest X-rays using Variational Data Imputation," no. August, 2020.

[88] K. He, G. Gkioxari, P. Dollár, and R. Girshick, "Mask R-CNN," *IEEE Trans. Pattern Anal. Mach. Intell.*, vol. 42, no. 2, pp. 386–397, Feb. 2020.

[89] J. Irvin *et al.*, "CheXpert: A large chest radiograph dataset with uncertainty labels and expert comparison," in *33rd AAAI Conference on Artificial Intelligence, AAAI 2019, 31st Innovative Applications of Artificial Intelligence Conference, IAAI 2019 and the 9th AAAI Symposium on Educational Advances in Artificial Intelligence, EAAI 2019*, 2019, pp. 590–597.

[90] Kaggle, "RSNA Pneumonia Detection Challenge | Kaggle," 2020. [Online]. Available: https://www.kaggle.com/c/rsna-pneumonia-detection-challenge. [Accessed: 23-Jan-2021].

[91]    Z. Yue, L. Ma, and R. Zhang, "Comparison and Validation of Deep Learning Models for the Diagnosis of Pneumonia," *Comput. Intell. Neurosci.*, vol. 2020, 2020.

**Appendix 1 - Technical aspects of LungX**

Turning to the more technical aspects of LungX, the solution is built on three different types of convolutional neural networks (CNNs). The first of the three models is a densenet 121 [86] that is able to detect the presence of the six different lung abnormalities as well as their location on a given X-ray.

The two additional models are used to calculate the severity score for COVID-19 patients. Only one of the six findings is related to COVID-19, and if this finding (edema/consolidation) is detected, the two additional models will calculate the severity score. A u-net [87] is used to segment the lungs into 12 pre-defined anatomical zones, while a mask-RCNN [88] is applied to segment the opacity in the lungs. When the outputs from the two models are mapped together, it is possible to calculate how many lung zones are affected by opacity. This score of how many of the 12 lung zones are affected is then used as the severity score indicating how badly the lungs are affected.

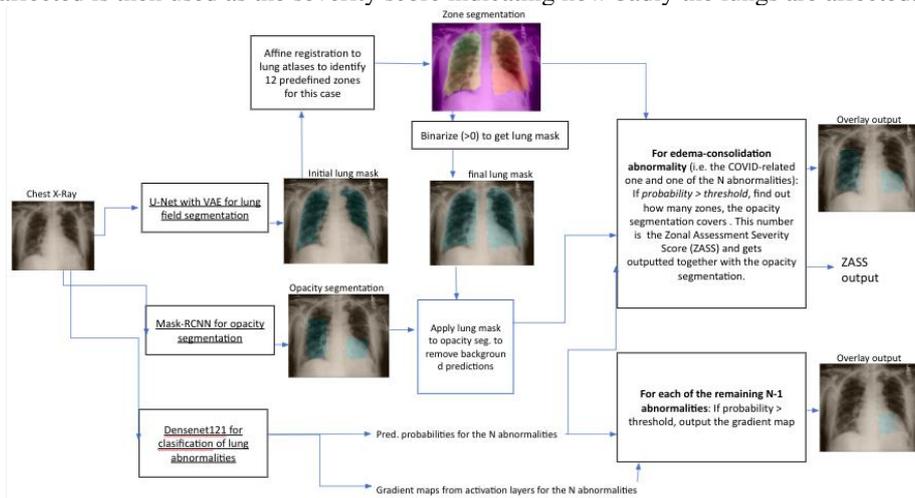

*Figure 7 Visualization of how the networks in LungX operate together*

The triage category indicates whether triage is low, medium, or high. The score is based on the severity score and the presence of any lung abnormalities the system is able to detect. It is configurable, meaning that the clinics using the system decide how the six abnormalities and the severity score for COVID-19 patients should each rank in relation to one another, as well as the level of triage. The highest triage category detected takes precedence.

The data foundation for training the developed model consists of two open-access data sources that are carefully joined to train the full model. Sources combined 112,120 frontal-view X-ray images of 30,805 unique patients from the Kaggle RSNA Pneumonia Detection Challenge in .PNG format and 224.316 radiographs from the CheXpert dataset from 65,240 patients who underwent a radiographic examination from Stanford University Medical Center [89]–[91]. However, none of these datasets include examples of COVID-19. COVID-19 examples were only provided from the hospital, in collaboration with the university project. 200 COVID patients tested positive for COVID-19 were run only on the models for prediction of disease progression.

| Box A | Box B |
|---|---|
| • Prediction of intubation<br>  • With Age, classification probabilities, ZASS – AUC: 0.70<br>  • ZASS LR: p=1e-8, OR=1.25<br>• Prediction of death<br>  • With Age, classification probabilities, ZASS – AUC: 0.80<br>  • ZASS LR: p=1e-4, OR=1.15<br>• Prediction of ICU<br>  • With Age, classification probabilities, ZASS – AUC: 0.71<br>  • ZASS LR: p=1e-9, OR=1.25 | • AUC<br>• Atelectasis – 0.70<br>• Normal – 0.94<br>• Lung lesion – 0.79<br>• Pneumothorax – 0.89<br>• Plueral effusion - 0.87<br>• Pnuemonia Opacity – 0.86 |

*Figure 1Performance metrics for LungX on 200 COVID-19 patients (box A) and on the CheXpert dataset (box B)*